\documentclass[12pt,letterpaper]{article}
\usepackage{epsfig,rotating,setspace,latexsym,amsmath,epsf,amssymb,amsfonts,bm,theorem,cite,caption,subcaption,enumerate,longtable,accents}
\usepackage{algorithm,algorithmic,graphicx,epsf,authblk,epstopdf,url,color,multirow}
\newtheorem{theorem}{Theorem}

\newtheorem{remark}{Remark}
\newtheorem{lemma}{Lemma}
\newenvironment{Proof}[1]{\medskip\par\noindent{\bf Proof:\,}\,#1}{{\mbox{\,$\blacksquare$}\par}}

\newcommand{\bg}{{\mathbf{G}}}
\newcommand{\bs}{{\mathbf{S}}}
\newcommand{\cq}{{\mathcal{Q}}}

\newcommand{\cu}{{\mathcal{U}}}
\newcommand{\cw}{{\mathcal{W}}}
\newcommand{\ct}{{\mathcal{T}}}
\newcommand{\mds}{{\text{\textbf{MDS}}}}

\allowdisplaybreaks

\begin{document}
	
\title{The Capacity of Private Information Retrieval from Byzantine and Colluding Databases\thanks{This work was supported by NSF Grants CNS 13-14733, CCF 14-22111, and CNS 15-26608.}}
	
\author{Karim Banawan \qquad Sennur Ulukus\\
	\normalsize Department of Electrical and Computer Engineering\\
	\normalsize University of Maryland, College Park, MD 20742 \\
	\normalsize {\it kbanawan@umd.edu} \qquad {\it ulukus@umd.edu}}
	
\maketitle
	
\vspace*{-0.8cm}

\begin{abstract}	
We consider the problem of single-round private information retrieval (PIR) from $N$ replicated databases. We consider the case when $B$ databases are outdated (unsynchronized), or even worse, adversarial (Byzantine), and therefore, can return incorrect answers. In the PIR problem with Byzantine databases (BPIR), a user wishes to retrieve a specific message from a set of $M$ messages with zero-error, irrespective of the actions performed by the Byzantine databases. We consider the $T$-privacy constraint in this paper, where any $T$ databases can collude, and exchange the queries submitted by the user. We derive the information-theoretic capacity of this problem, which is the maximum number of \emph{correct symbols} that can be retrieved privately (under the $T$-privacy constraint) for every symbol of the downloaded data. We determine the exact BPIR capacity to be $C=\frac{N-2B}{N}\cdot\frac{1-\frac{T}{N-2B}}{1-(\frac{T}{N-2B})^M}$, if $2B+T < N$. This capacity expression shows that the effect of Byzantine databases on the retrieval rate is equivalent to removing $2B$ databases from the system, with a penalty factor of $\frac{N-2B}{N}$, which signifies that even though the number of databases needed for PIR is effectively $N-2B$, the user still needs to access the entire $N$ databases. The result shows that for the unsynchronized PIR problem, if the user does not have any knowledge about the fraction of the messages that are mis-synchronized, the single-round capacity is the same as the BPIR capacity. Our achievable scheme extends the optimal achievable scheme for the robust PIR (RPIR) problem to correct the \emph{errors} introduced by the Byzantine databases as opposed to \emph{erasures} in the RPIR problem. Our converse proof uses the idea of the cut-set bound in the network coding problem against adversarial nodes.
\end{abstract}
	
\section{Introduction}
The problem of preserving the privacy of the contents downloaded from open-access databases has been a major area of research within the computer science community \cite{ChorPIR, PIRsurvey2004, cachin1999computationally, ostrovsky2007survey, yekhanin2010private}. Many practical applications are related to the private retrieval problem, such as: protecting the identity of stock market records reviewed by an investor, as showing interest in a specific record may undesirably affect its value; and protecting the nature of restricted content browsed by activists on the internet in oppressive regimes. In the seminal paper \cite{ChorPIR}, Chor et. al. introduced the problem of private information retrieval (PIR). In the classical PIR setting, a user wishes to retrieve a certain message (or file) out of $M$ distinct messages from $N$ non-colluding and replicated databases without leaking any information about the identity of the desired message. To that end, the user prepares $N$ queries, one for each database, in a single round, such that the queries do not reveal the user's interest in the desired message. Each database responds \emph{truthfully} with an answering string. The user needs to be able to reconstruct the entire message by decoding the answer strings from all databases. A straightforward solution for the PIR problem is for the user to download the entire database. This solution, however, is highly inefficient. The efficiency of PIR systems is assessed by the PIR rate, which is the ratio between the desired message size and the total downloaded symbols.

% PIR literature, computer science, information theory
The computer science formulation of the PIR problem assumes that the message is of length 1. The formulation considers optimizing two performance metrics, namely, the download cost, which is the sum of the lengths of the answer strings, and the upload cost, which is the sum of the lengths of the queries. Most of this work adopts computational guarantees as a privacy constraint, in which the databases cannot infer any information about the identity of the desired message unless they solve certain computationally hard problems\cite{yekhanin2010private, cachin1999computationally}.  Recently, the PIR problem is revisited by information theorists \cite{RamchandranPIR, YamamotoPIR, RazanPIR, JafarPIR}. The problem is re-formulated such that: the size of the message can be arbitrarily large, the upload cost is ignored, and privacy is guaranteed in the information-theoretic sense. This formulation gives rise to the PIR capacity notion, which is the supremum of PIR rates over all achievable retrieval schemes. In the pioneering paper \cite{JafarPIR}, Sun and Jafar determine the capacity of the classical PIR model, and propose a greedy algorithm which is based on three principles: message symmetry, database symmetry, and exploitation of side information through interference alignment as observed earlier in \cite{JafarPIRBlind}.

Several interesting extensions for the classical PIR problem are investigated following the information-theoretic reformulation in \cite{JafarPIR}, such as: PIR with $T$ colluding databases (TPIR) \cite{JafarColluding}, where the privacy constraint should be maintained against any $T$ databases; robust PIR (RPIR) \cite{JafarColluding}, where some databases fail to respond to the user; symmetric PIR (SPIR) \cite{symmetricPIR}, where the privacy of the remaining messages should be maintained against the user in addition to the usual user's privacy; MDS-coded PIR (CPIR) \cite{KarimCoded, RazanPIR}, where the contents of the databases are not replicated, but coded via an $(N,K)$ MDS code; multi-message PIR (MPIR) \cite{MPIRjournal}, where the user wishes to jointly retrieve $P$ messages; PIR under message size constraint $L$ (LPIR) \cite{arbmsgPIR}; multi-round PIR, where the queries are permitted to be a function of the answer strings collected in previous rounds \cite{MultiroundPIR}; MDS-coded symmetric PIR \cite{codedsymmetric}; MDS-coded PIR with colluding databases \cite{codedcolluded, codedcolludedJafar, codedcolludingZhang}, and its multi-message version~\cite{MPIRcodedcolludingZhang}.

% BPIR motivation, and details
A common assumption in these works is that the databases respond truthfully with the correct answer strings. Since the answers are correct, the user can use the undesired symbols downloaded from one database as side information at other databases. Furthermore, this enables the user to distribute the requests for the desired symbols among the $N$ databases. This poses an interesting question, how can we manage to reconstruct the desired message with no errors even if $B$ databases respond with incorrect answer strings? This question has practical implications. Returning to the examples presented earlier: The databases storing the stock market records may not be updated simultaneously, therefore some of the databases may store outdated versions of the messages and can introduce errors to the answering strings, which in turn leads to failure to reconstruct the desired message. This scenario is referred to in the literature as the \emph{unsynchronized PIR} problem \cite{unsynchonizedPIR}. For the oppressive regime example, some databases can be controlled by the regime, and these databases may return incorrect answer strings on purpose to confuse the user. This scenario is referred to in the literature as the \emph{PIR with adversarial databases} problem \cite{BiemelByzantine, OptimalRobust}. This motivates our interest in characterizing the exact capacity of the PIR problem with Byzantine databases (BPIR). In BPIR, there exist $B$ databases, which are called Byzantine databases, that respond with erroneous answer strings. The errors introduced by the Byzantine databases can be unintentional (as in the case of databases storing a different copy of the message set), or even worse, can be intentional (as in the case of maliciously controlled databases). In both cases, the user needs to be able to reconstruct the desired message with no error, irrespective of the actions performed by the Byzantine databases.

% BPIR literature
The BPIR problem was introduced in \cite{BiemelByzantine}. They propose a generic transformation from schemes of RPIR to robust protocols that tolerate Byzantine servers, and give an explicit Byzantine robust scheme when $B \leq T \leq \frac{N}{3}$.\cite{YangBPIR} presents a fault-tolerant PIR scheme that can cope with malicious failures for $B \leq T \leq \frac{N}{2}$. \cite{OptimalRobust} observes that allowing for list decoding instead of unique decoding enlarges the feasible set up to $B<N-T-1$. Their achievable scheme allows for a small failure probability. The scheme depends on Shamir's secret sharing algorithm \cite{ShamirSecretShare} and Guruswami-Sudan decoding algorithm \cite{SudanDecoding}. The unsynchronized PIR problem is investigated in \cite{unsynchonizedPIR}, where they propose a two-round retrieval scheme. The scheme returns the desired record by first identifying which records are mis-synchronized, and then by constructing a PIR scheme that avoids these problematic records.

% BPIR paper
In this paper, we consider the \emph{single-round} BPIR problem from $N$ replicated databases in the presence of $B$ Byzantine databases that can introduce errors to the returned answer strings. Other than the Byzantine databases, the remaining storage nodes store the exact copy of the message set which contains $M$ different messages, and respond truthfully with the correct answer strings. We consider the $T$-privacy constraint, which permits colluding between any $T$ databases to exchange the queries submitted by the user. Our goal is to characterize the single-round capacity of the BPIR problem under the zero-error reliability constraint and the $T$-privacy constraint. To that end, we propose an achievable scheme that is resilient to the worst-case errors that result from the Byzantine databases. Our achievable scheme extends the optimal scheme for the RPIR problem to correct the \emph{errors} resulted from the Byzantine databases, in contrast to the \emph{erasures} introduced by the unresponsive databases in RPIR. The new ingredients to the achievable scheme are: encoding the undesired symbols via a punctured MDS code, successive interference cancellation of the side information, and encoding the desired symbols by an outer-layer MDS code. For the converse, we extend the converse arguments developed for the network coding problem in \cite{TseByzantine} and distributed storage systems in \cite{SalimSecureDSS} to the PIR problem. This cut-set upper bound can be thought of as a network version of the Singleton bound \cite{SingletonBound}. The upper bound intuitively implies that a redundancy of $2B$ nodes is needed in order to mitigate the errors introduced by the $B$ Byzantine databases.

We determine the exact capacity of the BPIR problem to be $C=\frac{N-2B}{N}\cdot\frac{1-\frac{T}{N-2B}}{1-\left(\frac{T}{N-2B}\right)^M}$, if $2B+T<N$. The capacity expression shows the severe degradation of the retrieval rate due to the presence of Byzantine databases. The capacity expression is equivalent the TPIR capacity with $N-2B$ databases with a multiplicative factor of $\frac{N-2B}{N}$, which signifies the ignorance of the user as to which $N-2B$ databases are honest. Note that our Byzantine formulation includes the special case of the single-round unsynchronized PIR problem, if the user has no knowledge about the number of mis-synchronized messages, and only knows that the entirety of some $B$ databases may be unsynchronized. This formulation differs from the unsynchronized PIR setting in \cite{unsynchonizedPIR}, where a small number of records $S \ll M$ are mis-synchronized, and they allow for multi-round schemes. Under the assumptions of small number of mis-synchronized records and utilizing multiple rounds of querying (assuming no further mis-synchronization between the rounds) higher PIR rates may be achieved \cite{unsynchonizedPIR}. However, under our assumptions of up to the entire database being mis-synchronized and allowing only a single-round of querying, the single-round capacity of the unsynchronized PIR problem and the BPIR problem are the same.

\section{Problem Formulation}
Consider a single-round PIR setting with $N$ replicated databases storing $M$ messages (or files). The messages $\cw=\{W_1, \cdots, W_M\}$ are independent and uniformly distributed over a large enough finite field $\mathbb{F}_q$. Each message $W_i \in \mathbb{F}_q^L$ is a vector of length $L$ ($q$-ary symbols),
\begin{align}
H(W_i)&=L ,\quad i=1, \cdots, M \\
H(\cw)&=H(W_1,\cdots, W_M)=ML
\end{align}
Each database stores a copy from the complete set of messages $\cw$, i.e., this distributed storage system applies an $(N,1)$ repetition code \cite{KarimCoded}. Denote the contents of the $n$th database by $\Omega_n$. Ideally, $\Omega_n=\cw$ for all $n\in \{1, \cdots, N\}$.

In the PIR problem, a user wishes to retrieve a message $W_i \in \cw$ without revealing any information about the message index $i$. The user submits a single-round query $Q_n^{[i]}$ to the $n$th database. The user does not know the stored messages in advance, therefore, the message set $\cw$ and the queries are statistically independent,
\begin{align}
I\left(\cw;Q_{1:N}^{[i]}\right)=I\left(W_1,\cdots, W_M;Q_{1:N}^{[i]}\right)=0
\end{align}
where $Q_{1:N}^{[i]}=\{Q_1^{[i]},Q_2^{[i]},\cdots, Q_N^{[i]}\}$ is the set of all queries to the $N$ databases for message $i$.

Ideally, the classical PIR formulation assumes that all databases store the correct database contents (i.e., up-to-date contents), and respond truthfully with the correct answering strings $A_{1:N}^{[i]}=\{A_1^{[i]}, \cdots, A_N^{[i]}\}$. In the BPIR setting, on the other hand, there exists a set $\mathcal{B}$ of databases, that is unknown to the user, such that $|\mathcal{B}|=B$, which are called Byzantine databases. These databases can respond arbitrarily to the user by introducing errors to the answer strings $A_\mathcal{B}^{[i]}=\{A_j^{[i]}: j \in \mathcal{B}\}$, i.e.,
\begin{align}
H\left(A_n^{[i]}|Q_n^{[i]}, \cw\right)>0, \quad n \in \mathcal{B}, \: |\mathcal{B}|=B
\end{align}
We assume that these Byzantine databases can coordinate upon submitting the answers. In this paper, we do not assume a specific pattern to the errors. The remaining set of databases $\bar{\mathcal{B}}=\{1, \cdots, N\} \setminus \mathcal{B}$ respond truthfully to the user, i.e., the answer strings of $\bar{\mathcal{B}}$ are a deterministic function of the queries and the correct contents of the databases $\cw$,
\begin{align}\label{correct_answers}
H\left(A_n^{[i]}|Q_n^{[i]}, \cw\right)=0, \quad n \in \bar{\mathcal{B}}, \: |\bar{\mathcal{B}}|=N-B
\end{align}

We consider a $T$-privacy constraint as in the TPIR problem in \cite{JafarColluding}, where any $T$ databases can communicate and exchange the queries submitted by the user. To ensure the $T$-privacy constraint, the queries to any set $\mathcal{T} \subset \{1, \cdots, N\}$ of databases, such that $|\mathcal{T}|=T$, need to be statistically independent of the desired message index $i$, i.e.,
\begin{align}\label{T-privacy}
I\left(i;Q_\mathcal{T}^{[i]}\right)=0, \quad \text{for all} \: \mathcal{T}\subset \{1, \cdots, N\},\: |\mathcal{T}|=T
\end{align}
where $Q_\mathcal{T}^{[i]}$ are the queries submitted to the set $\ct$ of databases.

We remark here to differentiate the actions of \emph{colluding} between the databases which is done to figure out the desired message, and \emph{coordination} between the Byzantine databases which is done to introduce errors in the answer strings. In addition to the difference in their purposes, these two actions differ in the manner they are performed: colluding between any $T$ databases occurs upon receiving the queries from the user, while coordination between the $B$ Byzantine databases occurs upon submitting the answers to the user. We do not assume any specific relation between the $T$ colluding databases and the $B$ Byzantine databases.

The user should be able to reconstruct the desired message $W_i$, no matter what the Byzantine databases do, i.e., if there exists a set of databases $\bar{\mathcal{B}}$, that is unknown to the user, such that (\ref{correct_answers}) holds, then the reliability constraint is given by,
\begin{align}\label{reliability}
H(W_i|A_{1:N}^{[i]},Q_{1:N}^{[i]})=0, \quad \text{such that (\ref{correct_answers}) holds}
\end{align}

We define the \emph{resilient} PIR rate $R$ for the BPIR problem as the ratio between the message size $L$ and the total download cost under the reliability constraint in (\ref{reliability}) for any possible action of the Byzantine databases, and the $T$-privacy constraint in (\ref{T-privacy}), i.e.,
\begin{align}
R=\frac{L}{\sum_{n=1}^N H(A_n^{[i]})}
\end{align}
The capacity of BPIR is $C=\sup\: R$ over all possible single-round retrieval schemes.

In this paper, we follow the information-theoretic assumptions of large enough message size, large enough field size, and ignore the upload cost as in \cite{JafarColluding, KarimCoded, YamamotoPIR}. A formal treatment of the capacity under message size constraints can be found in \cite{arbmsgPIR}. The BPIR with colluding databases reduces to the TPIR problem in \cite{JafarColluding} if $B=0$.

Some scenarios that fit our formulation include:
\begin{itemize}
\item \emph{Unsynchronized setting \cite{unsynchonizedPIR}:} In this case, there exists a set $\mathcal{B}$ of databases, such that $|\mathcal{B}|=B$, in which they store different versions of the database contents, i.e.,
	\begin{align}
	\Omega_n \neq \cw,\quad  n \in \mathcal{B}, \: |\mathcal{B}|=B
	\end{align}
    Note that unlike \cite{unsynchonizedPIR}, we assume that the user has no knowledge about the fraction of the messages that are mis-synchronized. Hence, our achievable schemes must be resilient against the worst-case that the entirety of the database is mis-synchronized. Furthermore, the scheme in \cite{unsynchonizedPIR} is a two-round scheme, hence we cannot compare our rates with the rates in \cite{unsynchonizedPIR}; we consider only single-round schemes here.
\item \emph{Adversarial attacks \cite{BiemelByzantine, YangBPIR, OptimalRobust}:} In this case, the databases in $\mathcal{B}$ intend to preclude the retrieval process at the user by introducing a carefully-designed error sequence. This can be done by altering the contents of the databases to an erroneous version as in the unsynchronized setting; or by altering the answering strings themselves, i.e., the $n$th database returns the answer string $\tilde{A}_n^{[i]}$ such that,
	\begin{align}
	\tilde{A}_n^{[i]} \neq A_n^{[i]},\quad  n \in \mathcal{B}, \: |\mathcal{B}|=B
	\end{align}
    or by doing both.
\end{itemize}

\section{Main Result and Discussions}
The main result of this paper is to characterize the capacity of the BPIR problem under $T$-privacy constraint, where $B$ databases are adversarial (Byzantine) and can return malicious answers, and at the same time the privacy should be kept against any $T$ colluding databases.

\begin{theorem}\label{thm1}
For the single-round BPIR problem with $B$ Byzantine databases, and $T$ colluding databases, such that $2B+T<N$,  the capacity is given by,
\begin{align}
\label{result}	C &= \frac{N-2B}{N} \cdot \frac{1-\frac{T}{N-2B}}{1-\left(\frac{T}{N-2B}\right)^M} \\
	    &=\frac{N-2B}{N}\cdot\left(1+\frac{T}{N-2B}+\frac{T^2}{(N-2B)^2}+\cdots+\frac{T^{M-1}}{(N-2B)^{M-1}}\right)^{-1}
\end{align}
On the other hand, if $2B+1 \leq N \leq 2B+T$, then the user is forced to download the entire database from at least from $(2B+1)$ different databases, hence $C=\frac{1}{(2B+1)M}$, which is the trivial rate in the BPIR problem. Otherwise, the problem is infeasible and $C=0$.
\end{theorem}

The achievability proof for Theorem~\ref{thm1} is given in Section~\ref{Achievability}, and the converse proof is given in Section~\ref{converse}. We have a few remarks.

\begin{figure}[t]
	\centering
	\includegraphics[width=0.7\textwidth]{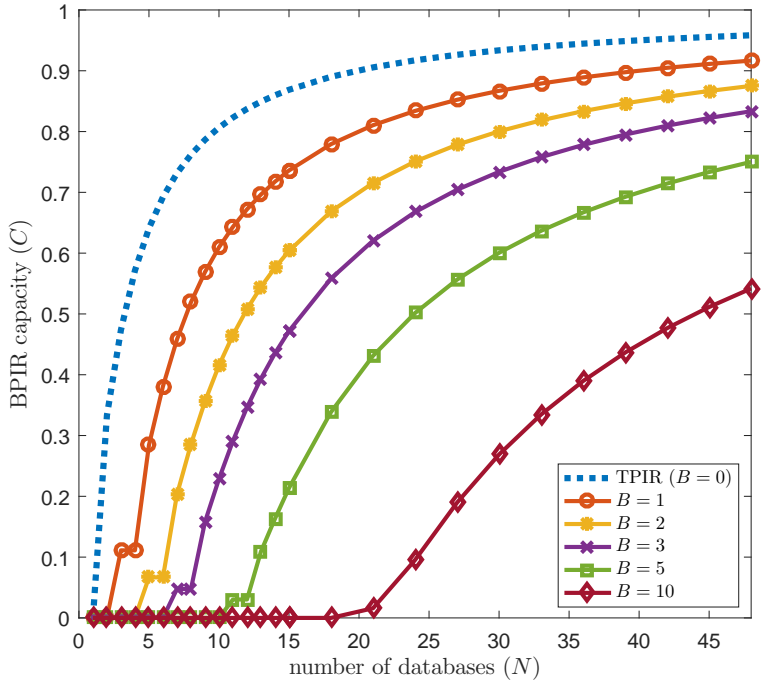}
	\caption{The effect of Byzantine databases on the BPIR capacity as a function of $N$ for fixed $T=2$, $M=3$.}
	\label{capacity_plots}
\end{figure}

\begin{remark}
The BPIR capacity in (\ref{result}) is the same as the capacity of PIR with $T$ colluding databases if the number of databases is $N-2B$ with a penalty factor of $\frac{N-2B}{N}$. This means that the harm introduced by the $B$ Byzantine databases is equivalent to removing a part from the storage system of size $2B$, but the user still needs to download from all $N$ databases, as it does not know which $N-2B$ databases are honest. This results in the penalty term $\frac{N-2B}{N}$. If $B=0$, the expression in (\ref{result}) reduces to
\begin{align}
C_{\text{colluded}} = \frac{1-\frac{T}{N}}{1-\left(\frac{T}{N}\right)^M}
\end{align}
which is the capacity expression in \cite{JafarColluding} as expected. Fig.~\ref{capacity_plots} shows the severe effect of the Byzantine databases on the retrieval rate for fixed $T=2$ and $M=3$ as a function of $N$.
\end{remark}

\begin{remark}
Comparing the BPIR capacity in Theorem~\ref{thm1} with the robust capacity $C_{\text{robust}}$ in \cite{JafarColluding}, where $U$ databases are merely unresponsive,
\begin{align}
C_{\text{robust}}=\frac{1-\frac{T}{N-U}}{1-\left(\frac{T}{N-U}\right)^M}
\end{align}
we note that the number of redundant databases, which are needed to correct the \emph{errors} introduced by the Byzantine databases, is \emph{twice} the number of redundant databases needed to correct the \emph{erasures} introduced in the case of unresponsive databases. We also note that the penalty factor is missing in the RPIR problem, since in the RPIR problem, the user does not get the chance to download from the unresponsive databases, in contrast to the BPIR problem, in which the user downloads answer strings from all databases. This is due to the fact that the user cannot identify the Byzantine databases before decoding the entire answer strings in the BPIR setting, while in the RPIR setting, the user identifies the unresponsive databases as they simply do not return answer strings.
\end{remark}

\begin{remark}
The trivial rate for the BPIR problem is $\frac{1}{(2B+1)M}$, which is much less than the trivial rate without the Byzantine databases, $\frac{1}{M}$. The reason for this is that the user cannot download the entire database only once in BPIR, but it must download $(2B+1)$ different copies of the database in order to decode the desired message via majority decoding. If $N<2B+1$, the capacity is $C=0$, as the Byzantine databases can always confuse the user to decode the desired message incorrectly.
\end{remark}

\begin{remark}
When the number of messages is large, i.e., as $M \rightarrow \infty$, the BPIR capacity $C \rightarrow (\frac{N-2B}{N})(1-\frac{T}{N-2B})=1-\frac{2B+T}{N}$, i.e., for large enough number of messages, the capacity expression acts as if there are no Byzantine databases and $2B+T$ databases are colluding.
\end{remark}

\begin{figure}[t]
	\centering
	\includegraphics[width=0.55\textwidth]{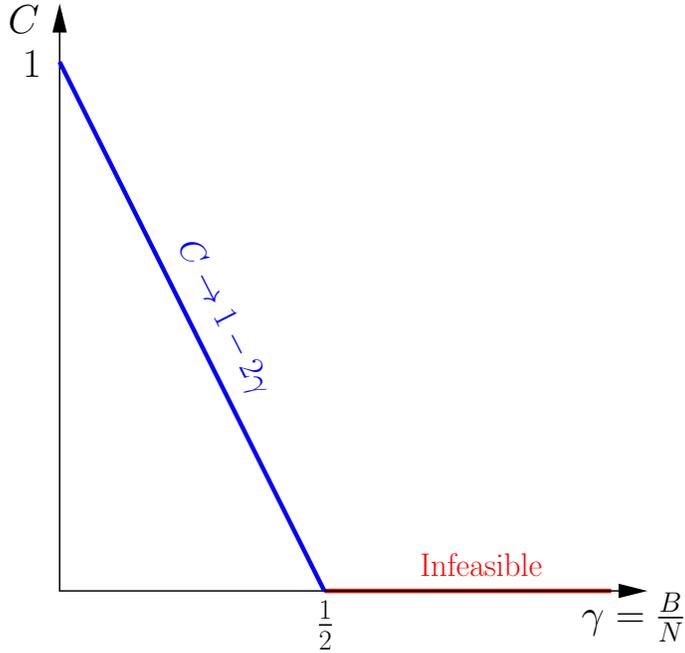}
	\caption{The asymptotic BPIR capacity $C$ as $N \rightarrow \infty$ as a function of $\gamma=\frac{B}{N}$.}
	\label{asymp_capacityN}
\end{figure}

\begin{remark}
If $T$ and $B$ are fixed and do not scale with $N$, i.e., $T=B=o(N)$, then the capacity is a strictly increasing function in $N$ and $C \rightarrow 1$ as $N \rightarrow \infty$. If the number of the Byzantine databases scales with $N$, i.e., $B=\gamma N$, where $\gamma \in \left[0,\frac{1}{2}(1-\frac{T}{N})\right)$, then $C\rightarrow 1-2\gamma$ as $N \rightarrow \infty$. If $2\gamma+\frac{1}{N} \leq 1 \leq 2\gamma+\frac{T}{N}$, then the only possible rate is the trivial rate $\frac{1}{(2B+1)M}$. As $N \rightarrow \infty$, then $\gamma \rightarrow \frac{1}{2}$, and $C \rightarrow 0$. This entails that the asymptotic behaviour of the BPIR capacity is a linear function with a slope of $-2$ as in Fig.~\ref{asymp_capacityN}, i.e., the asymptotic rate as $N \rightarrow \infty$ is decreased by \emph{twice} the ratio of the Byzantine databases. A similar behaviour is observed for secure distributed storage systems against Byzantine attacks in \cite{SalimSecureDSS}. The problem is infeasible if $\gamma>\frac{1}{2}$, i.e., $C=0$. This feasibility result conforms with the best result of a \emph{uniquely decodable} BPIR scheme in \cite{YangBPIR} which needs $B<\frac{N}{2}$.
\end{remark}

\begin{remark}
Surprisingly, our retrieval scheme in Section~\ref{Achievability} is a linear scheme in contrast to the network coding problem in \cite{TseByzantine} that states that linear coding schemes are not sufficient. We note that although the retrieval process is itself linear, the decoding process employs a successive interference cancellation decoder, which is non-linear.
\end{remark}

\begin{remark}
The capacity expression in Theorem~\ref{thm1} is also the capacity result for the unsychronized PIR problem \cite{unsynchonizedPIR}. This occurs under the restriction to single-round schemes and the assumption that the user only knows that there exist $B$ databases that are unsynchronized, but does not know the fraction of messages that are mis-synchronized. The achievability scheme in Section~\ref{Achievability} is a valid achievable scheme for the unsynchronized PIR problem, since the adversary in the Byzantine setting is stronger. For the converse proof, we restricted the actions of the adversarial databases to changing the contents of the stored messages, i.e., altering $\Omega_n$ from $\cw$ to $\tilde{\cw}$, which is the same setting as the unsynchronized PIR with no restriction on the fraction of messages that can be mis-synchronized.
\end{remark}

\section{Achievability Proof}\label{Achievability}
In this section, we present an achievable scheme that is resilient to the errors introduced by the Byzantine databases. The achievable scheme does not assume any specific error pattern. Hence, our achievable scheme enables \emph{correct decoding} of any desired message if any $B$ databases become outdated, or even worse, intentionally commit an adversarial attack to confuse the user. The achievable scheme generalizes the RPIR scheme presented in \cite{JafarColluding}. Our scheme has two new ingredients, namely, correcting errors in the side information using punctured MDS codes, and correcting errors in the desired message by an outer layer of MDS code. Error correction in both cases is performed via a nearest-codeword decoder.

\subsection{Preliminaries}
We start by presenting some preliminary results that will be needed. The following lemma states that if an MDS code is punctured by a puncture pattern whose length is smaller than the minimum distance of the original MDS code, then it remains an MDS code \cite{MDSpuncture}.

\begin{lemma}[MDS code puncturing \cite{MDSpuncture}]
	If $\mathcal{C}$ is an $(n,k)$ MDS code, then by puncturing the code by a sequence of length $z$, i.e., deleting a sequence of size $z$ from output codewords of $\mathcal{C}$, such that $z<n-k$, the resulting punctured code $\mathcal{C}_z$ is an $(n-z,k)$ MDS code.
\end{lemma}
The second lemma is regarding the statistical effect of operating on a random matrix by a deterministic full-rank matrix. The proof of this lemma can be found in \cite{JafarColluding}.

\begin{lemma}[Statistical effect of full-rank matrices \cite{JafarColluding}]\label{Lemma:full rank}
	Let $\bs_1,\bs_2,\cdots, \bs_M \in \mathbb{F}_q^{\alpha \times \alpha}$ be $M$ random matrices, drawn independently and uniformly from all $\alpha \times \alpha$ full-rank matrices over $\mathbb{F}_q$. Let $\bg_1, \bg_2, \cdots, \bg_M \in \mathbb{F}_q^{\beta \times \beta}$ be $M$ invertible square matrices of dimension $\beta \times \beta$ over $\mathbb{F}_q$. Let $\mathcal{I}_1, \cdots, \mathcal{I}_M \in \mathbb{N}^{\beta}$ be $M$ index vectors, each containing $\beta$ distinct indices from $\{1, \cdots, \alpha\}$, then
    \begin{align}
    \left\{\bg_1\bs_1(\mathcal{I}_1,:), \cdots, \bg_M\bs_M(\mathcal{I}_M,:)\right\} \sim \left\{(\bs_1([1:\beta],:), \cdots, \bs_M([1:\beta],:)\right\}
    \end{align}
    where $\sim$ denotes statistical equivalence, $\bs_i(\mathcal{I}_i,:),\, \bs_i([1:\beta],:)$ denote $\beta \times \alpha$ matrices with rows indexed by $\mathcal{I}_i$ and $\{1,2, \cdots, \beta\}$, respectively.
\end{lemma}
The next lemma summarizes the code capabilities of handling errors and erasures for linear block codes \cite[Theorem~1.7]{RothCoding}.

\begin{lemma}[Code capabilities \cite{RothCoding}]\label{Rothlemma}
	Let $\mathcal{C}$ be an $[n,k,d]$ linear block code over $\mathbb{F}_q$. Let $\rho$ be the number of erasures introduced by the channel. Let $\tau \in \mathbb{N}$, such that $2\tau+\rho \leq d-1$, then there exists a nearest-codeword decoder that recovers all errors and erasures if the number or errors (excluding erasures) is $\tau$ or less.
\end{lemma}
Lemma~\ref{Rothlemma} implies that in the case of no erasures, the maximum number of errors $\tau \leq \left\lfloor\frac{d-1}{2}\right\rfloor$.

\subsection{Motivating Example: $M=2$ Messages, $N=5$, $T=2$, $B=1$ Databases} \label{sec:mot}
Assume without loss of generality that $W_1$ is the desired message. Let $a_i$ and $b_i$ be the $i$th symbol mixture of messages $W_1$ and $W_2$, respectively. The specific construction of these mixtures will be presented shortly. We begin the retrieval process by downloading $T^{M-1}=2$ symbols from $W_1$, which are $a_1, a_2$ as in \cite{JafarColluding}. By \emph{message symmetry}, we download $2$ symbols from $W_2$, which are $b_1,b_2$. By \emph{database symmetry}, we download $2$ symbols from $W_1$ and $2$ symbols from $W_2$ from all other databases.

Now, we want to generate the maximum number of side information equations in order to maximize the retrieval rate. From Lemma~\ref{Rothlemma}, we see that the number of errors that can be corrected increases with $d$. We know that MDS codes meet the Singleton bound \cite{SingletonBound} with equality, hence encoding both desired and undesired messages by MDS codes is desirable. In addition, Lemma~\ref{Rothlemma} implies a \emph{doubling effect}, which suggests that in order to correct the errors introduced by the Byzantine database, we should effectively consider $N-2B=3$ honest databases. Consequently, considering any $3$ databases, the number of undesired symbols is $6$. We note that any $T=2$ of them can collude, therefore, we are left with $2$ undesired symbols that can be used to generate side information among the $2$ colluding databases. Hence, each database should get $1$ side information equation $b_{[11:15]}$. These side-information symbols can be added to new desired symbols $a_{[11:15]}$. The complete query structure is shown in Table~\ref{Table(2,5,2,1)}.

\begin{table}[h]
	\centering
	\caption{The query table for the case $M=2$, $N=5$, $T=2$, $B=1$.}
	\label{Table(2,5,2,1)}
	\begin{tabular}{|c|c|c|c|c|}
		\hline
		DB 1 & DB 2 & DB 3 & DB 4 & DB 5 \\
		\hline
		$a_1$ & $a_3$  & $a_5$ & $a_7$ & $a_9$ \\
		$a_2$ & $a_4$  & $a_6$ & $a_8$ & $a_{10}$ \\
		$b_1$ & $b_3$  & $b_5$ & $b_7$ & $b_9$ \\
		$b_2$ & $b_4$  & $b_6$ & $b_8$ & $b_{10}$ \\
		\hline
		$a_{11}+b_{11}$ & $a_{12}+b_{12}$  & $a_{13}+b_{13}$ & $a_{14}+b_{14}$ & $a_{15}+b_{15}$ \\
		\hline
	\end{tabular}
\end{table}

Now, we specialize the query structure in Table~\ref{Table(2,5,2,1)}, and identify the specific construction of the mixtures $a_{[1:15]}$ and $b_{[1:15]}$. For the desired message $W_1$, considering any $N-2B=3$ honest databases, we see $9$ distinct symbols. Therefore, the length of $W_1$ is $L=9$, and we use $\bs_1$, which is a $9 \times 9$ random mixing matrix picked uniformly from the full-rank matrices over $\mathbb{F}_q^{9 \times 9}$. These $9$ mixed symbols are further mapped to $a_{[1:15]}$ by a $(15,9)$ MDS code generator matrix $\mds_{15 \times 9}$, therefore,
\begin{align}
a_{[1:15]}=\mds_{15 \times 9} \bs_1 W_1
\end{align}
For the undesired message $W_2$, considering again any $N-2B=3$ honest databases, we have $6$ individual symbols from $W_2$ in round 1. We should be able to reconstruct the side information equations $b_{[11:15]}$ in round 2 from any $6$ individual symbols, hence we get $6$ random symbols from $W_2$. This can be done by considering the first $6$ rows of the random mixing matrix $\bs_2 \in \mathbb{F}_q^{9 \times 9}$. These randomly mixed symbols are further mapped to $b_{[1:15]}$ via and MDS code with generator matrix $\mds_{15 \times 6}$, i.e.,
\begin{align}
b_{[1:15]}=\mds_{15 \times 6} \bs_2([1:6],:) W_2
\end{align}

To see the decodability: the worst-case scenario is that the Byzantine database commits errors in all the symbols returned to the user. This means that the database commits $2$ errors in the individual symbols from $W_1$, $2$ errors in the individual symbols from $W_2$, and $1$ extra error in the sum of $a+b$.

Consider the codeword $b_{[1:10]}$: this codeword belongs to $(15,6)$ MDS code with a sequence of length $z=5$ removed. Hence, this codeword belongs to $(10,6)$ punctured MDS code. Since $z=5<15-6=9$, the $(10,6)$ punctured MDS code is still an MDS code. Denote the minimum distance of the $(10,6)$ punctured MDS code that results in $b_{[1:10]}$ by $d_p^b$. Then, $d_p^b=10-6+1=5$. Consequently, from Lemma~\ref{Rothlemma}, the $(10,6)$ punctured MDS code can tolerate errors up to $\tau_b$, such that
\begin{align}
\tau_b \leq \left\lfloor\frac{d_p^b-1}{2}\right\rfloor=2
\end{align}
Therefore, this code can correct all errors that can be introduced to the individual undesired symbols $b_{[1:10]}$. Let $b_{[1:10]}^*$ be the correct codeword of $b_{[1:10]}$. Choose any $6$ symbols from $b_{[1:10]}^*$. Now, since $\mds_{15 \times 6}$ matrix has the property that any $6 \times 6$ matrix is an invertible matrix, then from any $6$ symbols from $b_{[1:10]}^*$, the \emph{correct side information} equations $b_{[11:15]}^*$ are determined and canceled from the sums of $a$ and $b$ in round 2.

For the desired message $W_1$: after removing the interference from $W_2$, we are left with $\tilde{a}_{[1:15]}$. Note that this is not exactly $a_{[1:15]}$, because we canceled the correct side information and not $b_{[1:15]}$. However, the total errors in $\tilde{a}_{[1:15]}$ still is upper bounded by $3$, since $\tilde{a}_{[1:15]}$ can differ from $a_{[1:15]}$ only in the positions that correspond to Byzantine databases. The desired message $W_1$ is coded via $(15,9)$ MDS code. Then, the minimum distance for this code is $d^a=15-9+1=7$. Consequently, this code can tolerate errors up to $\tau_a$, such that
\begin{align}
\tau_a \leq \left\lfloor\frac{d^a-1}{2}\right\rfloor=3
\end{align}
Hence, all the errors in $\tilde{a}_{[1:15]}$ can be corrected, and we can obtain true $a_{[1:15]}^*$. Consider the first $9$ symbols from $a_{[1:15]}^*$, without loss of generality, then
\begin{align}
W_1=(\mds_{15 \times 9} ([1:9],:) \bs_1)^{-1} a_{[1:9]}^*
\end{align}
since $\mds_{15 \times 9} ([1:9],:) \bs_1$ is a $9 \times 9$ invertible matrix.

Therefore, despite Byzantine behaviour of $B=1$ database, we decode the desired message correctly. In addition, our achievable scheme can identify the Byzantine database as does the scheme in \cite{unsynchonizedPIR} by comparing $a_{[1:10]}^*$ with $a_{[1:10]}$, and $b_{[1:10]}^*$ with $b_{[1:10]}$ and see which database has introduced errors.

To see the privacy: we note that from any $T=2$ databases, our achievable scheme collects $6$ symbols from $a_{[1:15]}$ and $6$ symbols from $b_{[1:15]}$ indexed by $\mathcal{I}$ such that $|\mathcal{I}|=6$. For the undesired message, we collect $b_\mathcal{I}$,
\begin{align}
b_\mathcal{I}   &=\mds_{15 \times 6}(\mathcal{I},:) \bs_2([1:6],:) W_2 \\
\label{ex11}    &\sim \bs_2([1:6],:) W_2
\end{align}
where (\ref{ex11}) follows from Lemma~\ref{Lemma:full rank} as any $6 \times 6$ matrix in $\mds_{15 \times 6}$ matrix is full-rank. Therefore, the symbols $b_\mathcal{I}$ are independent and uniformly distributed. For $a_\mathcal{I}$, we have
\begin{align}
a_\mathcal{I}&=\mds_{15 \times 9}(\mathcal{I},:) \bs_1 W_1 \\
\label{ex12} &=\Psi_{6 \times 9} W_1
\end{align}
where $\Psi=\mds_{15 \times 9}(\mathcal{I},:) \bs_1$ is a full row-rank matrix as any $6$ rows in $\mds_{15 \times 9}$ are linearly independent. Consequently, the symbols $a_\mathcal{I}$ are also independent and uniformly distributed, and  $a_\mathcal{I} \sim b_\mathcal{I}$ for every $2$ databases, where $\sim$  means that the involved random vectors are statistically identical. Thus, the proposed scheme is $2$-private; that is, despite colluding behaviour of $T=2$ databases, we have privacy.

Finally, the achievable resilient retrieval rate is $R=\frac{9}{25}=\frac{N-2B}{N}\cdot\frac{1-\frac{T}{N-2B}}{1-\left(\frac{T}{N-2B}\right)^M}=C$. In comparison, the trivial rate for this system is $\frac{1}{(2B+1)M}=\frac{1}{6}$, as the user must download the entire database from $3$ different databases for correct decoding.

\subsection{General Achievable Scheme}
The general achievable scheme is performed in $M$ rounds. The $i$th round includes  all the $\binom{M}{i}$ combinations of the sums of any $i$ messages. The scheme requires $L=(N-2B)^{M}$. The construction resembles the optimal scheme for RPIR in \cite{JafarColluding}. The new key ingredient in our achievable scheme is the decoding procedure, which includes correcting the undesired symbols by punctured MDS codes, successive interference cancellation to cancel the interfering messages, and correcting the errors in the desired message by an outer layer MDS code.

\subsubsection{General Description for the Scheme}
\begin{enumerate}
\item \emph{Initialization:} The scheme starts with downloading $T^{M-1}$ mixed symbols from the desired message from the first database. The specific construction of the mixture will be specified shortly. The scheme sets the round index $i=1$.

\item \emph{Message symmetry:} To satisfy the privacy constraint, the user downloads the same number of mixed symbols from the undesired messages with all the possible combinations, i.e., in the $i$th round, the user downloads $\binom{M-1}{i} (N-2B-T)^{i-1} T^{M-i}$ mixed symbols from the remaining $M-1$ messages. The specific construction of the undesired mixture will be specified shortly.
	
\item \emph{Database symmetry:} The user repeats the same steps at all the databases. Specifically, the user downloads $\binom{M-1}{i-1} (N-2B-T)^{i-1} T^{M-i}$ equations in the form of a desired message mixture symbol and $i-1$ mixed symbols from the undesired messages, and $\binom{M-1}{i} (N-2B-T)^{i-1} T^{M-i}$ mixed symbols from the undesired messages only, from each database.
	
\item \emph{Exploiting side information:} The specific construction of the undesired mixtures should be done such that in the $(i+1)$th round, the user should be able to generate $\frac{N-2B-T}{T}$ side information equations for each undesired symbol in the $i$th round. This fraction is a consequence of considering $\tilde{N}=N-2B$ honest databases only, and dividing the undesired symbols from the $\tilde{N}-T$ databases among the $T$ colluding databases. The side information generated is added to a new mixed symbol from the desired message.
	
\item Repeat steps 2, 3, 4 after setting $i=i+1$ until $i=M-1$.
\end{enumerate}

\subsubsection{Specific Construction of the Symbol Mixtures}
Let $W_m \in \mathbb{F}_q^{(N-2B)^M}, \: m \in \{1, \cdots, M\}$ be the message vectors, and $\bs_m, \: m \in  \{1, \cdots, M\}$ be random mixing matrices picked independently and uniformly from the full-rank matrices in $\mathbb{F}_q^{(N-2B)^M \times (N-2B)^M}$. From the general description of the scheme, we note that at the $i$th round, the user downloads all possible combinations of the sums of any $i$ messages. In the following specific construction, we enumerate all the sets that contain a symbol from the desired message and assign them labels $\mathcal{L}_1, \cdots, \mathcal{L}_\delta$. For each undesired message, we further enumerate also all the sets that contain symbols from this undesired message and do not include any desired symbols and assign them labels $\mathcal{K}_1, \cdots, \mathcal{K}_\Delta$. These sets construct the undesired symbol mixtures and the corresponding side information.

For the desired message: Assume that the desired message is $W_\ell$. Let $\delta$ be the number of the distinct subsets of $\{1, \cdots, M\}$ that contain $\ell$, then $\delta=2^{M-1}$. Let $\mathcal{L}_i, \: i \in \{1, \cdots, \delta\}$ be the $i$th subset that contains $\ell$. Assume without loss of generality, that these sets are arranged in ascending order in the sizes of the sets $|\mathcal{L}_i|$. According to this order, we note that $\mathcal{L}_1=\{\ell\}$ and belongs to round 1. Round 2 contains sets $\mathcal{L}_2, \cdots, \mathcal{L}_{\binom{M-1}{1}+1}$, and so on. Let $X^{[\ell]} \in \mathbb{F}_q^{N(N-2B)^M}$ be the vector of mixtures that should be obtained from the desired message $W_\ell$. Divide $X^{[\ell]}$ into $\delta$ partitions denoted by $x_{\mathcal{L}_i}^{[\ell]}$, each corresponds to a distinct set $\mathcal{L}_i$. Now, encode the desired message by a $\left(N(N-2B)^{M-1},(N-2B)^{M} \right)$ MDS code as,
\begin{align}
X^{[\ell]}=\begin{bmatrix}
x_{\mathcal{L}_1}^{[\ell]} \\ x_{\mathcal{L}_2}^{[\ell]} \\ \vdots \\ x_{\mathcal{L}_\delta}^{[\ell]}
\end{bmatrix}
=\mds_{N(N-2B)^{M-1} \times (N-2B)^M} \bs_\ell W_l
\end{align}
where $x_{\mathcal{L}_i}^{[\ell]}$ is a vector of length $N(N-2B-T)^{|\mathcal{L}_i|-1} T^{M-|\mathcal{L}_i|}$ in $\mathbb{F}_q$.

For any other undesired message: Consider the undesired message $W_k$, $k \in \{1, \cdots, M\} \setminus \{\ell\}$. Let $\Delta=2^{M-2}$ be the number of distinct subsets that contain $k$ and do not contain $\ell$. Let $\mathcal{K}_i$, $i \in \{1, \cdots, \Delta\}$ be the $i$th subset that contains $k$ and does not contain $\ell$ with indices in ascending order in the size of set $|\mathcal{K}_i|$. Define $u_{\mathcal{K}_i}^{[k]}$ to be the undesired symbol mixtures in the $|\mathcal{K}_i|$th round corresponding to message $k$ among the $\mathcal{K}_i$ set. Define $\sigma_{\mathcal{K}_i}^{[k]}$ to be the side information symbols from message $k$ among the $\mathcal{K}_i$ subset of undesired messages. These side information equations are added to a desired message symbol in the $(|\mathcal{K}_i|+1)$th round. For each subset $\mathcal{K}_i$, the undesired symbols and side information symbols are related via,
\begin{align}
\begin{bmatrix}
u_{\mathcal{K}_i}^{[k]} \\ \sigma_{\mathcal{K}_i}^{[k]}
\end{bmatrix}
=\mds_{\frac{N}{T} \alpha_i \times \alpha_i} \bs_k \left(\left[\sum_{j=1}^{i-1}\alpha_j+1: \sum_{j=1}^{i}\alpha_j\right],:\right) W_k
\end{align}
where $\alpha_i=(N-2B)(N-2B-T)^{|\mathcal{K}_i|-1} T^{M-|\mathcal{K}_i|}$, $u_{\mathcal{K}_i}^{[k]}$ is a vector of length $\frac{N}{N-2B} \alpha_i$, and $\sigma_{\mathcal{K}_i}^{[k]}$ is a vector of length $\frac{N-2B-T}{T}\cdot\frac{N}{N-2B}  \alpha_i$. This implies that the side information $\sigma_{\mathcal{K}_i}^{[k]}$ in the $(|\mathcal{K}_i|+1)$th round is completely determined by $u_{\mathcal{K}_i}^{[k]}$ in the $|\mathcal{K}_i|$th round. We note that these choices of the dimensions ensure that the same number of desired and undesired symbols exist in the $|\mathcal{K}_i|$th round, and they are both equal to $N(N-2B-T)^{|\mathcal{K}_i|-1}T^{M-|\mathcal{K}_i|}$. We further note that the $\frac{N-2B-T}{T}$ factor in the length of $\sigma_{\mathcal{K}_i}^{[k]}$, implies that we generate $\frac{N-2B-T}{T}$ side information symbols for each undesired symbol. We note that the same MDS matrix is used for all messages $k \neq \ell$ that belong to the same subset $\mathcal{K}_i$. This is critical to enable \emph{interference alignment}, and \emph{joint error correction}. Let $X^{[k]} \in \mathbb{F}_q^{N(N-2B)^{M-1}}$ be the vector of mixtures corresponding to message $k \neq \ell$. Then,
\begin{align}
X^{[k]}\!=\!\begin{bmatrix}
u_{\mathcal{K}_1}^{[k]} \\ \sigma_{\mathcal{K}_1}^{[k]} \\ u_{\mathcal{K}_2}^{[k]} \\ \sigma_{\mathcal{K}_2}^{[k]} \\ \vdots \\ u_{\mathcal{K}_\Delta}^{[k]} \\ \sigma_{\mathcal{K}_\Delta}^{[k]}
\end{bmatrix}
\!=\!\begin{bmatrix}
\mds_{\frac{N}{T}\alpha_1 \times \alpha_1} \! &\! \mathbf{0}\! &\! \cdots \!&\! \mathbf{0} \\
\mathbf{0} \!&\!\mds_{\frac{N}{T}\alpha_2 \times \alpha_2} \! & \! \cdots \!&\!\mathbf{0}  \\
\vdots & \vdots & \vdots & \vdots \\
\mathbf{0} \!&\! \mathbf{0}  \!&\! \mathbf{0} \!&\! \mds_{\frac{N}{T}\alpha_\Delta \times \alpha_\Delta}
\end{bmatrix}\! \bs_k ([1:T(N\!-\!2B)^{M-1}],:) W_k
\end{align}

Now, we are ready to specify the queries. For every non-empty set $\mathcal{M} \subseteq \{1, \cdots, M\}$, define $\cq_\mathcal{M}^{[\ell]}$ to be all queries related to set $\mathcal{M}$,
\begin{align}
\cq_\mathcal{M}^{[\ell]}=
 \left\{
 \begin{array}{ll}
 x_{\mathcal{L}_1}^{[\ell]},  & \mathcal{M}=\mathcal{L}_1=\{\ell\} \\
 x_{\mathcal{L}_j}^{[\ell]}+\sum_{k \in \mathcal{K}_i} \sigma_{\mathcal{K}_i}^{[k]} \, & \exists i,j: \mathcal{M}=\mathcal{K}_i \cup \{\ell\}=\mathcal{L}_j \\
 \sum_{k \in \mathcal{K}_i} u_{\mathcal{K}_i}^{[k]} \, & \exists i: \mathcal{M}=\mathcal{K}_i
 \end{array}
 \right.
\end{align}
We distribute the queries randomly and evenly among the $N$ databases for each subset $\mathcal{M}$, and the construction is now complete.

\subsection{Decodability, Privacy, and the Achievable Rate}
First, we show how the decoding is performed. The first step is to correct the errors in the undesired symbols in the $\mathcal{K}_i$ set in the $|\mathcal{K}_i|$th round, so that we can generate the correct side information in the $(|\mathcal{K}_i|+1)$th round. Consider again the encoding,
\begin{align}
\begin{bmatrix}
u_{\mathcal{K}_i}^{[k]} \\ \sigma_{\mathcal{K}_i}^{[k]}
\end{bmatrix}
=\mds_{\frac{N}{T} \alpha_i \times \alpha_i} \bs_k \left(\mathcal{J}_i,:\right) W_k
\end{align}
where $\mathcal{J}_i=\left[\sum_{j=1}^{i-1}\alpha_j+1: \sum_{j=1}^{i}\alpha_j\right]$. Since the sum of linear codes is also a linear code, for the every set $\mathcal{K}_i$, $i \in \{1, \cdots, \Delta\}$, we have
 \begin{align}
 \begin{bmatrix}
  \sum_{k \in \mathcal{K}_i}u_{\mathcal{K}_i}^{[k]} \\ \sum_{k \in \mathcal{K}_i}\sigma_{\mathcal{K}_i}^{[k]}
 \end{bmatrix}
 =\mds_{\frac{N}{T} \alpha_i \times \alpha_i} \sum_{k \in \mathcal{K}_i}\bs_k \left(\mathcal{J}_i,:\right) W_k
 \end{align}
This enables \emph{joint error correction} on the aligned sum. The minimum distance of this MDS code is $d^{\mathcal{K}_i}=\frac{N}{T} \alpha_i - \alpha_i+1=\frac{N-T}{T} \alpha_i+1$.

Now, in the $|\mathcal{K}_i|$th round, the user downloads $\sum_{k \in \mathcal{K}_i}u_{\mathcal{K}_i}^{[k]}$ which is a vector of length $\frac{N}{N-2B} \alpha_i$ from all databases. The vector $\sum_{k \in \mathcal{K}_i}u_{\mathcal{K}_i}^{[k]}$ belongs to $\left(\frac{N}{N-2B} \alpha_i, \alpha_i\right)$ punctured MDS code with a puncturing sequence corresponding to the side information symbols, i.e., with a puncturing sequence of length $z=|\sigma_{\mathcal{K}_i}^{[k]}|=\frac{N-2B-T}{T} \cdot \frac{N}{N-2B}\alpha_i$. Therefore,
\begin{align}
d^{\mathcal{K}_i}-z-1&=\frac{N-T}{T} \alpha_i -\frac{N-2B-T}{T} \cdot  \frac{N}{N-2B} \alpha_i \\
&=\frac{2B}{N-2B} \alpha_i \\
&=2B(N-2B-T)^{|\mathcal{K}_i|-1} T^{M-|\mathcal{K}_i|} >0
\end{align}
Thus, the $\left(\frac{N}{N-2B} \alpha_i, \alpha_i\right)$ punctured MDS code remains an MDS code with a minimum distance $d^{u_i}$, such that
\begin{align}
d^{u_i}&=\frac{N}{N-2B} \alpha_i-\alpha_i+1\\
       &=\frac{2B}{N-2B} \alpha_i +1
\end{align}
Hence, the punctured code can correct upto $\tau_{u_i} $ errors, such that
\begin{align}
\tau_{u_i} \leq \left\lfloor\frac{d^{u_i}-1}{2}\right\rfloor=\frac{B}{N-2B} \alpha_i
\end{align}
Each database contributes $\frac{1}{N-2B} \alpha_i$ symbols from $\sum_{k \in \mathcal{K}_i}u_{\mathcal{K}_i}^{[k]}$, hence the Byzantine databases can introduce at most $\frac{B}{N-2B} \alpha_i$ errors. Consequently, the punctured MDS code can correct all errors in $\sum_{k \in \mathcal{K}_i}u_{\mathcal{K}_i}^{[k]}$. This results in a corrected undesired message vector $\left(\sum_{k \in \mathcal{K}_i}u_{\mathcal{K}_i}^{[k]}\right)^*$. Choose any $\alpha_i$ symbols from $\left(\sum_{k \in \mathcal{K}_i}u_{\mathcal{K}_i}^{[k]}\right)^*$. By the MDS property of the $(\frac{N}{T} \alpha_i, \alpha_i)$ MDS code, any $\alpha_i \times \alpha_i$ submatrix is invertible, hence a correct version of the side information vector, which is used in the $(|\mathcal{K}_i|+1)$th round, can be generated. Denote this correct version by $\left(\sum_{k \in \mathcal{K}_i}\sigma_{\mathcal{K}_i}^{[k]}\right)^*$.

Now, we cancel the correct side information successively from each set $\mathcal{K}_i$. Note that the successive correction of side information gives rise to non-linearity in the decoding. After interference cancellation, we are left with $\tilde{X}^{[\ell]}$, which is not exactly $X^{[\ell]}$, as we cancelled the correct side information from the sum and not the side information provided by the Byzantine databases. This is not a problem, because $\tilde{X}^{[\ell]}$ and $X^{[\ell]}$ differ in codeword positions if and only if these positions belong to the Byzantine databases, hence the worst-case number of errors in $\tilde{X}^{[\ell]}$ cannot increase. The desired message is encoded by $(N(N-2B)^{M-1}, (N-2B)^M)$ MDS code with minimum distance $d^{x}$, such that
\begin{align}
d^{x}&=N(N-2B)^{M-1}-(N-2B)^M+1 \\
     &=2B(N-2B)^{M-1}+1
\end{align}
Each database returns $(N-2B)^{M-1}$ symbols from the desired message. The $B$ Byzantine databases can at most introduce $B(N-2B)^{M-1}$ errors. The outer MDS code can correct up to $\tau_x$ errors, such that
\begin{align}
\tau_{x} \leq \left\lfloor\frac{d^{x}-1}{2}\right\rfloor=B(N-2B)^{M-1}
\end{align}
Thus, the user can correct all the errors introduced by the Byzantine databases to get a correct vector $\left(X^{[\ell]}\right)^* \in \mathbb{F}_q^{N(N-2B)^{M-1}}$. Consider any $(N-2B)^M$ symbols from $\left(X^{[\ell]}\right)^*$. Denote these symbols by $x^*_\ell$, and index them by $\mathcal{I}_x$. Then, the user can decode $W_\ell$ with zero error via
\begin{align}
W_\ell=(\mds_{N(N-2B)^{M-1} \times (N-2B)^M}(\mathcal{I}_x,:)\bs_1)^{-1}x^*_\ell
\end{align}
This is true as matrix $\mds_{N(N-2B)^{M-1} \times (N-2B)^M}(\mathcal{I}_x,:)\bs_1$ is invertible by the MDS property.

In addition, the user can identify the Byzantine databases by comparing the correct versions of the undesired symbols at each cancellation step $(\sum_{k \in \mathcal{K}_i}u_{\mathcal{K}_i}^{[k]})^*$, and the desired symbols $\left(X^{[\ell]}\right)^*$ by their counterparts from the retrieval process. Any change between the correct vector and the retrieved vector implies that this database is a Byzantine database (or unsynchronized). The user can expurgate the  malicious nodes in this case as in \cite{SalimSecureDSS,unsynchonizedPIR,TseByzantine}.

Next, we show how the privacy is achieved. The queries for any $T$ colluding databases are comprised of $T(N-2B)^{M-1}$ mixed symbols from each message $W_i$, $i \in \{1, \cdots, M\}$. Let these symbols be indexed by $\mathcal{I}$. Denote the $k$th message symbols by $x_\mathcal{I}^{[k]}$. For the desired symbols, we have
\begin{align}
x_\mathcal{I}^{[\ell]}=\mds_{N(N-2B)^{M-1} \times (N-2B)^M} (\mathcal{I},:) \bs_\ell W_l
\end{align}
Since $|\mathcal{I}|=T(N-2B)^{M-1} < (N-2B)^M$ as $2B+T<N$ by construction, and due to the MDS property, the symbols $x_\mathcal{I}^{[\ell]}$ have full-rank. Hence, they are independent and uniformly distributed. Furthermore, for any undesired message $W_k, \: k \neq \ell$, we have,
\begin{align}
x_\mathcal{I}^{[k]}=\underbrace{\begin{bmatrix}
\mds_{\frac{N}{T}\alpha_1 \times \alpha_1}(\mathcal{I}_1,:) \! &\! \cdots \!&\! \mathbf{0} \\
\mathbf{0}  \! & \! \cdots \!&\!\mathbf{0}  \\
\vdots  & \vdots & \vdots \\
\mathbf{0}  \!&\! \mathbf{0} \!&\! \mds_{\frac{N}{T}\alpha_\Delta \times \alpha_\Delta}(\mathcal{I}_\Delta,:)
\end{bmatrix}}_{\Phi}\! \bs_k ([1:T(N\!-\!2B)^{M-1}],:) W_k
\end{align}
where $\mathcal{I}=\bigcup_{j=1}^\Delta \mathcal{I}_j$, and $|\mathcal{I}_j|=\alpha_j$. Therefore, each submatrix in $\Phi$ is an $\alpha_i \times \alpha_i$ invertible matrix by the MDS property. Hence, $\Phi$ is also an invertible matrix because it is a block-diagonal matrix. By Lemma~\ref{Lemma:full rank}, we have
\begin{align}
x_\mathcal{I}^{[k]} \sim \bs_k ([1:T(N\!-\!2B)^{M-1}],:) W_k
\end{align}
Thus, symbols $x_\mathcal{I}^{[k]}$ are independent and uniformly distributed, and the privacy is guaranteed.

We next calculate the achievable resilient rate. We note that the scheme operates in $M$ rounds. At the $i$th round, the scheme downloads $\binom{M-1}{i-1}(N-2B-T)^{i-1} T^{M-i}$ equations in the form of one desired symbol added to $i-1$ symbols from the undesired messages, and $\binom{M-1}{i} (N-2B-T)^{i-1} T^{M-i}$ undesired symbols only. Then, the total download in the $i$th round is $\binom{M}{i} (N-2B-T)^{i-1} T^{M-i}$ from each database, i.e., the total download of the scheme, $D$, is $D=N \sum_{i=1}^M \binom{M}{i} (N-2B-T)^{i-1} T^{M-i}$. The scheme decodes correctly the desired message, which has length $L=(N-2B)^M$. Thus, the resilient retrieval rate is,
\begin{align}
R   &=\frac{L}{D} \\
    &=\frac{(N-2B)^M}{N\sum_{i=1}^M \binom{M}{i} (N-2B-T)^{i-1} T^{M-i}} \\
    &=\frac{N-2B}{N}\cdot\frac{(N-2B)^{M-1}}{\sum_{i=1}^M \binom{M}{i} (N-2B-T)^{i-1} T^{M-i}} \\
    &=\frac{N-2B}{N}\cdot \frac{(N-2B)^{M-1}}{\frac{1}{N-2B-T} \sum_{i=1}^M \binom{M}{i} (N-2B-T)^{i} T^{M-i}} \\
    &=\frac{N-2B}{N}\cdot \frac{(N-2B)^{M-1}}{\frac{1}{N-2B-T} \left((N-2B)^M-T^M\right)} \\
    &=\frac{N-2B}{N}\cdot \frac{(N-2B)^M-T(N-2B)^{M-1}}{(N-2B)^M-T^M} \\
    &=\frac{N-2B}{N} \cdot \frac{1-\frac{T}{N-2B}}{1-\left(\frac{T}{N-2B}\right)^M}
\end{align}
which is the expression in Theorem~\ref{thm1}. We have additional some about the achievable scheme.

\begin{remark}
We note that our achievable scheme is capable of identifying the Byzantine databases by observing discrepancies between the corrected codewords of desired and undesired messages and their counterparts from the retrieval process. Therefore, if multiple-rounds are allowed in the achievable scheme, we can remove the databases that introduce errors at each retrieval round, and achieve larger retrieval rates in future rounds. For instance, assume that $\tilde{B} \leq B$ databases commit errors and are identified to be Byzantine in the $k$th retrieval round, then removing these databases from the system and downloading only from the remaining $(N-\tilde{B})$ databases, we can achieve the following retrieval rate in the $(k+1)$th round
\begin{align}
R^{(k+1)} &=\frac{N-\tilde{B}-2(B-\tilde{B})}{N-\tilde{B}}\cdot \frac{1-\frac{T}{N-\tilde{B}-2(B-\tilde{B})}}{1-(\frac{T}{N-\tilde{B}-2(B-\tilde{B})})^M}\\
	            &=\frac{N+\tilde{B}-2B}{N-\tilde{B}}\cdot \frac{1-\frac{T}{N+\tilde{B}-2B}}{1-(\frac{T}{N+\tilde{B}-2B})^M}
\end{align}
In particular, if all $B$ Byzantine databases act maliciously in the $k$th retrieval round and get identified, i.e., $\tilde{B}=B$, then we can achieve the following retrieval rate in the $(k+1)$th round
\begin{align}
R^{(k+1)}=\frac{1-\frac{T}{N-B}}{1-(\frac{T}{N-B})^M}
\end{align}
which is the retrieval rate if $B$ databases are just unresponsive.
\end{remark}

\begin{remark}
Our achievable scheme can be seamlessly extended to the case of BPIR with $U$ unresponsive databases (as in the case of RPIR \cite{JafarColluding}) -- also known in the literature as $T$-private $B$-Byzantine $(N-U)$-out-of-$N$ PIR as in \cite{OptimalRobust}. The construction of the achievable scheme can be done by replacing every $N-2B$ with $N-2B-U$ in the general achievable scheme. Using Lemma~\ref{Rothlemma}, that states that correct decoding is possible if $2\tau+\rho \leq d-1$, and considering the effect of the unresponsive databases as erasures, i.e., via $\rho$, the decodability holds for the BPIR problem with unresponsive databases. The retrieval rate in this case is,
\begin{align}
	R=\frac{N-2B-U}{N-U}\cdot\frac{1-\frac{T}{N-2B-U}}{1-(\frac{T}{N-2B-U})^M} \label{rateBPIRunresponsive}
\end{align}
The retrieval expression is the same as the BPIR capacity in (\ref{result}) if the number of databases is $N-U$. This in turn implies that the expression in (\ref{rateBPIRunresponsive}) is the capacity of the BPIR problem with unresponsive databases. The details of the construction and the analysis are omitted to avoid repetition.
\end{remark}

\subsection{Further Examples}
In this section, we present some further simple examples with tractable parameters of $M$, $N$, $T$, $B$ for better understanding of the achievable scheme. Here, we use increased number of messages ($M=3$) and databases ($N=6$) compared to the selections $M=2$, $N=5$ in the motivating example in Section~\ref{sec:mot}. In the following two subsections, we choose $T=1$, $B=2$ and $T=2$, $B=1$, respectively, to show the different effects of colluding and Byzantine behavior. We assume without loss of generality that the desired message is $W_1$.

\subsubsection{$M=3$ Messages, $N=6$, $T=1$, $B=2$ Databases}
We denote the mixed symbols of messages $W_1, W_2,W_3$ by $a,b,c$, respectively. In this example $L=(N-2B)^M=8$, hence we use $8 \times 8$ random mixing matrices denoted by $\bs_1,\bs_2,\bs_3$. We have $\mathcal{L}_1=\{1\}, \mathcal{L}_2=\{1,2\}, \mathcal{L}_3=\{1,3\}, \mathcal{L}_4=\{1,2,3\}$. Also, for the undesired message $W_2$, we have $\mathcal{K}_1=\{2\}, \mathcal{K}_2=\{2,3\}$, and similarly for $W_3$. The scheme starts with downloading $T^{M-1}=1$ symbol from each message from each database. Therefore, in round 1, the scheme downloads $x_{\mathcal{L}_1}^{[1]}=a_{[1:6]}$,  $u_{\mathcal{K}_1}^{[2]}=b_{[1:6]}$, and $u_{\mathcal{K}_1}^{[3]}=c_{[1:6]}$; see Table~\ref{Table(3,6,1,2)}. For every undesired symbol in round 1, we generate $\frac{N-2B-T}{T}=1$ side information symbols to be used in round 2. The scheme constructs the side information symbols $\sigma_{\mathcal{K}_1}^{[2]}=b_{[7:12]}$ based on the downloaded symbols $b_{[1:6]}$, and similarly for $\sigma_{\mathcal{K}_1}^{[3]}=c_{[7:12]}$. Round 2 contains all combinations of the sums of $2$ messages. Round 2 adds one new symbol from the desired message with one symbol of the generated side information from $b,c$. This results in the sums $x_{\mathcal{L}_2}^{[1]}+\sigma_{\mathcal{K}_1}^{[2]}=a_{[7:12]}+b_{[7:12]}$, and the sums $x_{\mathcal{L}_3}^{[1]}+\sigma_{\mathcal{K}_1}^{[3]}=a_{[13:18]}+c_{[7:12]}$. By message symmetry, we must include the undesired symbol sum $\sum_{k \in \mathcal{K}_2}u_{\mathcal{K}_2}^{[k]}=b_{[13:18]}+c_{[13:18]}$; see Table~\ref{Table(3,6,1,2)}. We note that these undesired information equation is in the form of \emph{aligned sums}. The undesired symbols in round 2 generate the side information equations $\sum_{k \in \mathcal{K}_2}\sigma_{\mathcal{K}_2}^{[k]}=b_{[19:24]}+c_{[19:24]}$. These side information equations are added to new symbols from the desired message to have $x_{\mathcal{L}_4}^{[1]}+\sum_{k \in \mathcal{K}_2}\sigma_{\mathcal{K}_2}^{[k]}=a_{[19:24]}+b_{[19:24]}+c_{[19:24]}$. The query table is shown in Table~\ref{Table(3,6,1,2)}.

\begin{table}[h]
	\centering
	\caption{The query table for the case $M=3$, $N=6$, $T=1$, $B=2$.}
	\label{Table(3,6,1,2)}
	\begin{tabular}{|c|c|c|c|c|c|}
		\hline
		DB 1 & DB 2 & DB 3 & DB 4 & DB 5 & DB 6 \\
		\hline
		$a_1$ & $a_2$  & $a_3$ & $a_4$ & $a_5$ & $a_6$ \\
		$b_1$ & $b_2$  & $b_3$ & $b_4$ & $b_5$ & $b_6$\\
		$c_1$ & $c_2$  & $c_3$ & $c_4$ & $c_5$ & $c_6$ \\
		\hline
		$a_7+b_7$ & $a_8+b_8$  & $a_9+b_9$ & $a_{10}+b_{10}$ & $a_{11}+b_{11}$ & $a_{12}+b_{12}$ \\
		$a_{13}+c_7$ & $a_{14}+c_8$  & $a_{15}+c_9$ & $a_{16}+c_{10}$ & $a_{17}+c_{11}$ & $a_{18}+c_{12}$ \\
		$b_{13}+c_{13}$ & $b_{14}+c_{14}$  & $b_{15}+c_{15}$ & $b_{16}+c_{16}$ & $b_{17}+c_{17}$ & $b_{18}+c_{18}$ \\
		\hline
		$a_{19}\!+\!b_{19}\!+\!c_{19}$ & $a_{20}\!+\!b_{20}\!+\!c_{20}$  & $a_{21}\!+\!b_{21}\!+\!c_{21}$ & $a_{22}\!+\!b_{22}\!+\!c_{22}$ & $a_{23}\!+\!b_{23}\!+\!c_{23}$ & $a_{24}\!+\!b_{24}\!+\!c_{24}$ \\
		\hline
	\end{tabular}
\end{table}

The specific construction of the symbol mixtures are,
\begin{align}
a_{[1:24]}&=\mds_{24 \times 8} \bs_1 W_1 \\
b_{[1:24]}&= \begin{bmatrix}u_{\mathcal{K}_1}^{[2]} \\\sigma_{\mathcal{K}_1}^{[2]} \\ u_{\mathcal{K}_2}^{[2]} \\ \sigma_{\mathcal{K}_2}^{[2]}  \end{bmatrix}
=\begin{bmatrix}
\mds_{12 \times 2} & \mathbf{0} \\
\mathbf{0} & \mds_{12 \times 2}
\end{bmatrix} \bs_2 ([1:4],:) W_2 \\
c_{[1:24]}&= \begin{bmatrix}u_{\mathcal{K}_1}^{[3]} \\\sigma_{\mathcal{K}_1}^{[3]} \\ u_{\mathcal{K}_2}^{[3]} \\ \sigma_{\mathcal{K}_2}^{[3]}  \end{bmatrix}
=\begin{bmatrix}
\mds_{12 \times 2} & \mathbf{0} \\
\mathbf{0} & \mds_{12 \times 2}
\end{bmatrix} \bs_3 ([1:4],:) W_3
\end{align}

For the decodability, we note that $B=2$ Byzantine databases can introduce at most $2$ errors in $b_{[1:6]}$, $2$ errors in $c_{[1:6]}$, $2$ errors in $b_{[13:18]}+c_{[13:18]}$, and $8$ errors in $a_{[1:24]}$. We note that $b_{[1:6]}$ is encoded via $(6,2)$ punctured MDS code, which still is an MDS code because $z=6<12-2=10$. The $(6,2)$ punctured MDS code can correct errors up to $\lfloor \frac{6-2}{2}\rfloor=2$ errors. Then, the $2$ errors in $b_{[1:6]}$ can be corrected. The same argument holds for $c_{[1:6]}$. For $b_{[13:18]}+c_{[13:18]}$, since the same generator matrix is used for $b_{[13:18]}$, $c_{[13:18]}$, and because of the linearity of the code, the \emph{aligned sum} is a codeword from $(6,2)$ punctured MDS code as well. Thus, we can correct all the errors in the aligned sum $b_{[13:18]}+c_{[13:18]}$. Knowing the correct undesired symbols results in decoding the correct side information symbols $b_{[7:12]}$, $c_{[7:12]}$ and $b_{[19:24]}+c_{[19:24]}$, respectively, by the MDS property. Cancelling these side information from the answer strings, we are left with $\tilde{a}_{[1:24]}$, which are coded with an outer $(24,8)$ MDS code, which is capable of correcting $\lfloor\frac{24-8}{2}\rfloor=8$ errors. Hence, the user can correct all the errors introduced by the Byzantine databases and $W_1$ is decodable.

For the privacy, from any individual database, the user asks for $4$ mixed symbols from each message. Because of the MDS property, the symbols from all messages are full-rank, and hence they are independent and uniformly distributed. Thus, the scheme is private.

The resilient achievable rate is $R=\frac{8}{42}=\frac{4}{21}=\frac{1}{3} \cdot \frac{4}{7} = \frac{N-2B}{N} \cdot \frac{1-\frac{T}{N-2B}}{1-\left(\frac{T}{N-2B}\right)^M}=C$.

\subsubsection{$M=3$ Messages, $N=6$, $T=2$, $B=1$ Databases}
In this case $L=(N-2B)^M=64$, and we use random mixing matrices $\bs_1,\bs_2,\bs_3$ of size $64 \times 64$. The scheme starts by downloading $T^{M-1}=4$ symbols from each message from each database, namely, $a_{[1:24]}$, $b_{[1:24]}$, $c_{[1:24]}$; see Table~\ref{Table(3,6,2,1)}. The undesired symbols from $b_{[1:24]}$ and $c_{[1:24]}$ create $\frac{N-2B-T}{T}=1$ side information symbol for each undesired symbol in a single database. Therefore, the scheme generates the side information $b_{[25:48]}$, $c_{[25:48]}$. In round 2, these side information are added to $a_{[25:48]}, a_{[49:72]}$, respectively. Round 2 concludes by applying message symmetry, and downloads $b_{[49:72]}+c_{[49:72]}$. These undesired symbols produce $b_{[73:96]}+c_{[73:96]}$ as side information symbols for round 3. The query table is shown in Table~\ref{Table(3,6,2,1)}.

The specific construction of the symbol mixtures are,
\begin{align}
a_{[1:96]}&=\mds_{96 \times 64} \bs_1 W_1 \\
b_{[1:96]}&=
\begin{bmatrix}
\mds_{48 \times 16} & \mathbf{0} \\
\mathbf{0} & \mds_{48 \times 16}
\end{bmatrix} \bs_2 ([1:32],:) W_2 \\
c_{[1:96]}&=
\begin{bmatrix}
\mds_{48 \times 16} & \mathbf{0} \\
\mathbf{0} & \mds_{48 \times 16}
\end{bmatrix} \bs_3 ([1:32],:) W_3
\end{align}

For the decodability, the Byzantine database can commit $4$ errors in $b_{[1:24]}$, $4$ errors in $c_{[1:24]}$, $4$ errors in $b_{[49:72]}+c_{[49:72]}$, and $16$ errors in $a_{[1:96]}$. All layers of the undesired symbols are encoded via $(24,16)$ punctured MDS code, which is still MDS code, and can correct up to $\lfloor\frac{24-16}{2}\rfloor=4$ errors. Therefore, all the undesired symbols can be corrected, which in turn generate the correct side information in all layers. By canceling the side information, we are left with $\tilde{a}_{[1:96]}$, which is encoded by $(96,64)$ outer MDS code. This code can correct up to $\lfloor\frac{96-64}{2} \rfloor=16$ errors. Hence, the user can decode $W_1$ reliably.

For the privacy, from any $2$ databases, the user asks for $16$ symbols from each message. By the MDS property and Lemma~\ref{Lemma:full rank}, all these symbols are full-rank, and hence they are independent and uniformly distributed. Therefore, the scheme is 2-private.

The resilient achievable rate is $R=\frac{64}{168}=\frac{8}{21}=\frac{4}{6}\cdot \frac{4}{7}=\frac{N-2B}{N}\cdot\frac{1-\frac{T}{N-2B}}{1-\left(\frac{T}{N-2B}\right)^M}=C$.

Note that, for the same $M$, $N$, the achievable rate with $T=1$, $B=2$ in the previous subsection, $\frac{4}{21}$, is smaller than the achievable rate with $T=2$, $B=1$ in this subsection, $\frac{8}{21}$, which signifies that Byzantine behavior is a more severe adversarial behavior to cope with compared to colluding behavior.

\begin{table}
	\centering
	\caption{The query table for the case $M=3$, $N=6$, $T=2$, $B=1$.}
	\label{Table(3,6,2,1)}
	\begin{tabular}{|c|c|c|c|c|c|}
		\hline
		DB 1 & DB 2 & DB 3 & DB 4 & DB 5 & DB 6 \\
		\hline
		$\!a_1,a_2,a_3,a_4\!$ & $\!a_5,a_6,a_7,a_8\!$  & $\!a_9,a_{10},a_{11},a_{12}\!$ & $\!\!a_{13},a_{14},a_{15},a_{16}\!\!$ &$\!\!a_{17},a_{18},a_{19},a_{20}\!\!$ & $\!\!a_{21},a_{22},a_{23},a_{24}\!\!$ \\
		$b_1,b_2,b_3,b_4$ & $b_5,b_6,b_7,b_8$  & $b_9,b_{10},b_{11},b_{12}$ & $b_{13},b_{14},b_{15},b_{16}$ &$b_{17},b_{18},b_{19},b_{20}$ & $b_{21},b_{22},b_{23},b_{24}$ \\
		$c_1,c_2,c_3,c_4$ & $c_5,c_6,c_7,c_8$  & $c_9,c_{10},c_{11},c_{12}$ & $c_{13},c_{14},c_{15},c_{16}$ &$c_{17},c_{18},c_{19},c_{20}$ & $c_{21},c_{22},c_{23},c_{24}$ \\
		\hline
		$a_{25}+b_{25}$ & $a_{29}+b_{29}$  & $a_{33}+b_{33}$ & $a_{37}+b_{37}$ & $a_{41}+b_{41}$ & $a_{45}+b_{45}$ \\
		
		$a_{26}+b_{26}$ &  $a_{30}+b_{30}$ & $a_{34}+b_{34}$ & $a_{38}+b_{38}$& $a_{42}+b_{42}$ & $a_{46}+b_{30}$ \\
		
		$a_{27}+b_{27}$ &  $a_{31}+b_{31}$ & $a_{35}+b_{35}$ & $a_{39}+b_{39}$& $a_{43}+b_{43}$ & $a_{47}+b_{47}$ \\
		
		$a_{28}+b_{28}$ &  $a_{32}+b_{32}$ & $a_{36}+b_{36}$ & $a_{40}+b_{40}$& $a_{44}+b_{44}$ & $a_{48}+b_{48}$ \\

		$a_{49}+c_{25}$ & $a_{53}+c_{29}$  & $a_{57}+c_{33}$ & $a_{61}+c_{37}$ & $a_{65}+c_{41}$ & $a_{69}+c_{45}$ \\
		
		$a_{50}+c_{26}$ &  $a_{54}+c_{30}$ & $a_{58}+c_{34}$ & $a_{62}+c_{38}$& $a_{66}+c_{42}$ & $a_{70}+c_{30}$ \\
		
		$a_{51}+c_{27}$ &  $a_{55}+c_{31}$ & $a_{59}+c_{35}$ & $a_{63}+c_{39}$& $a_{67}+c_{43}$ & $a_{71}+c_{47}$ \\
		
		$a_{52}+c_{28}$ &  $a_{56}+c_{32}$ & $a_{60}+c_{36}$ & $a_{64}+c_{40}$& $a_{68}+c_{44}$ & $a_{72}+c_{48}$ \\

		$b_{49}+c_{49}$ & $b_{53}+c_{53}$  & $b_{57}+c_{57}$ & $b_{61}+c_{61}$ & $b_{65}+c_{65}$ & $b_{69}+c_{69}$ \\
		
		$b_{50}+c_{50}$ &  $b_{54}+c_{54}$ & $b_{58}+c_{58}$ & $b_{62}+c_{62}$& $b_{66}+c_{66}$ & $b_{70}+c_{70}$ \\
		
		$b_{51}+c_{51}$ &  $b_{55}+c_{55}$ & $b_{59}+c_{59}$ & $b_{63}+c_{63}$& $b_{67}+c_{67}$ & $b_{71}+c_{71}$ \\
		
		$b_{52}+c_{52}$ &  $b_{56}+c_{56}$ & $b_{60}+c_{60}$ & $b_{64}+c_{64}$& $b_{68}+c_{68}$ & $b_{72}+c_{72}$ \\

		\hline
		$a_{73}\!+\!b_{73}\!+\!c_{73}$ & $a_{77}\!+\!b_{77}\!+\!c_{20}$  & $a_{81}\!+\!b_{81}\!+\!c_{81}$ & $a_{85}\!+\!b_{85}\!+\!c_{85}$ & $a_{89}\!+\!b_{89}\!+\!c_{89}$ & $a_{93}\!+\!b_{93}\!+\!c_{93}$ \\
		
		$a_{74}\!+\!b_{74}\!+\!c_{74}$ & $a_{78}\!+\!b_{78}\!+\!c_{78}$  & $a_{82}\!+\!b_{82}\!+\!c_{82}$ & $a_{86}\!+\!b_{86}\!+\!c_{86}$ & $a_{90}\!+\!b_{90}\!+\!c_{90}$ & $a_{94}\!+\!b_{94}\!+\!c_{94}$ \\
		
		$a_{75}\!+\!b_{75}\!+\!c_{75}$ & $a_{79}\!+\!b_{79}\!+\!c_{79}$  & $a_{83}\!+\!b_{83}\!+\!c_{83}$ & $a_{87}\!+\!b_{87}\!+\!c_{87}$ & $a_{91}\!+\!b_{91}\!+\!c_{91}$ & $a_{95}\!+\!b_{95}\!+\!c_{95}$ \\
		
		$a_{76}\!+\!b_{76}\!+\!c_{76}$ & $a_{80}\!+\!b_{80}\!+\!c_{80}$  & $a_{84}\!+\!b_{84}\!+\!c_{84}$ & $a_{88}\!+\!b_{88}\!+\!c_{89}$ & $a_{92}\!+\!b_{92}\!+\!c_{92}$ & $a_{96}\!+\!b_{96}\!+\!c_{96}$ \\
		\hline
	\end{tabular}
\end{table}

\section{Converse Proof}\label{converse}
In this section, we develop an upper bound for the BPIR problem. We adapt the cut-set upper bound proof in \cite{TseByzantine,SalimSecureDSS} to the PIR setting. The upper bound can be thought of as a network version of the Singleton bound \cite{SingletonBound}. The upper bound intuitively asserts that the effect of the Byzantine databases on the retrieval rate is harmful as if $2B$ databases are removed from the retrieval process, but the user still needs to access them. The settings of PIR and network coding problem in \cite{TseByzantine} share that they are both planar networks, and they both lack backward edges, as we consider here a single-round retrieval, and hence the answer strings from the honest databases are not affected by the answers of the Byzantine databases. However, some technical differences arise in the PIR setting:
\begin{enumerate}
\item Unlike the adversarial nodes in \cite{TseByzantine,SalimSecureDSS}, the Byzantine databases in PIR are not fully omniscient, since they do not know which message the user wishes to retrieve (by definition of PIR). Consequently, we assume in the following that the Byzantine databases alter the contents of the entire database.
	
\item In the PIR setting, the user does not know the entire codebook in advance, in contrast to the network coding problem in \cite{TseByzantine}.
\end{enumerate}
For sake of deriving an upper bound, we make the following simplifications:
\begin{enumerate}
\item We assume that the actions of the Byzantine databases are restricted to altering the contents of the entire database, i.e., the $n$th Byzantine database changes its contents $\Omega_n$ from $\mathcal{W}$ to $\mathcal{\tilde{W}}$, where $\mathcal{\tilde{W}} \neq \mathcal{W}$. This restriction is valid from the converse point of view, since it potentially results in a weaker adversary, which in turn results in a higher rate. Note that, in this sense the Byzantine databases are reduced to being unsynchronized databases (with unknown number of mis-synchronized messages).
	
\item We further restrict the answering string from the $n$th database to be a deterministic function $f_n (\cdot)$, i.e., $A_n^{[i]}=f_n(\Omega_n,Q_n^{[i]})$, of the altered database $\Omega_n$. This restriction also limits the capabilities of the Byzantine databases. This results in a further upper bound on rate. Since we restrict the actions of the Byzantine databases to altering $\Omega_n$ only, we signify this dependence on $\Omega_n$ by writing the answering string $A_n^{[i]}$ as $A_n^{[i]}(\Omega_n)$.
	
\item We can assume that the retrieval scheme is symmetric. This is without loss of generality, since any asymmetric PIR scheme can be made symmetric by proper time sharing without changing the retrieval rate \cite{JafarPIR, KarimCoded,MPIRjournal}, i.e.,
	\begin{align}
	H(A_1^{[i]}|\cq)=	H(A_2^{[i]}|\cq)= \cdots=	H(A_N^{[i]}|\cq)
	\end{align}
    This assumption remains true in the BPIR problem, because if the $n$th Byzantine database returned an answering string which has $H(A_n^{[i]}|\cq) \neq H(A_j^{[i]}|\cq)$ for some honest database $j$, i.e., the answering string has a different length as a response to a symmetric retrieval scheme, this database will be identified as a Byzantine database. Hence, the errors introduced by the Byzantine databases can be mitigated and these databases will be removed from the system afterwards. In addition, the restrictions in assumptions 1 and 2 above imply that the Byzantine databases answer truthfully to the queries based on their own (altered) $\Omega_n$. Therefore, the lengths of the answer strings will be symmetric in response to a symmetric scheme.
\end{enumerate}
The main argument of the converse proof is summarized in the following lemma.

\begin{lemma}\label{lemma:uniqueness}
	Fix a set of honest databases $\cu \subset \{1, \cdots, N\}$ such that $|\cu|=N-2B$, and $\Omega_n=\cw$, for every $n \in \cu$. Then, for correct decoding of $W_i$, the answer strings $A_\cu^{[i]}(\cw)$ is unique for every realization of $\cw$, i.e., there cannot exist two realizations of the message set $\cw, \tilde{\cw}$, such that $\cw \neq \tilde{\cw}$, and $A_\cu^{[i]}(\cw)=A_\cu^{[i]}(\tilde{\cw})$.
\end{lemma}
We have this following remark about Lemma~\ref{lemma:uniqueness} first, before we give its proof next.
\begin{remark}
Lemma~\ref{lemma:uniqueness} implies that the answer strings from any $N-2B$ honest databases are enough to reconstruct the desired message, since every realization of the message set produces different answering strings from any $N-2B$ databases. This argument was previously used by \cite[Theorem~1]{TseByzantine} and \cite[Theorem~6]{SalimSecureDSS}, as they show that the capacity of the adversarial network coding problem and the adversarial distributed storage problem, respectively, is upper bounded by the capacity of the edges of any cut in the network after removing $2B$ edges from this cut. These edges correspond to the set $\mathcal{U}$ in our problem. The proof in \cite{TseByzantine, SalimSecureDSS} relies on the fact that in the presence of an adversary controlling $B$ nodes, and for any distinct messages $w_1 \neq w_2$, a necessary condition for the receiver to not make a decoding error is to have $X_\mathcal{U}(w_1) \neq X_\mathcal{U} (w_2)$.
\end{remark}

\begin{Proof}
Divide the set $\bar{\cu}=\{1, \cdots, N\}\setminus \cu$ into two sets $\mathcal{B}_1$, $\mathcal{B}_2$ such that $|\mathcal{B}_1|=|\mathcal{B}_2|=B$. In the BPIR problem, we must guarantee correct decoding if the Byzantine databases are any subset $\mathcal{B} \subset \{1, \cdots, N\}$, such that $|\mathcal{B}|=B$, in particular, if the Byzantine databases are either $\mathcal{B}_1$ or $\mathcal{B}_2$.

Now, assume for sake of contradiction, that there exists a valid retrieval scheme that achieves correct decoding of $W_i$, and there exist two realizations of the message set $\cw,\, \tilde{\cw}$ such that $\cw \neq \tilde{\cw}$, and
\begin{align}
A_\cu^{[i]}(\cw)=A_\cu^{[i]}(\tilde{\cw})
\end{align}
Two scenarios can arise:
\begin{enumerate}
\item The true realization of the database contents is $\cw$. In this case, if the adversarial nodes are the databases indexed by $\mathcal{B}_2$, and they flip their contents $\Omega_{\mathcal{B}_2}$ into $\tilde{\cw}$, the user collects the answer strings $\left(A_{\mathcal{B}_1}^{[i]}(\cw),A_{\mathcal{B}_2}^{[i]}(\tilde{\cw}),A_\cu^{[i]}(\cw)\right)$.
	
\item The true realization of the database contents is $\tilde{\cw}$. In this case, if the adversarial nodes are the databases indexed by $\mathcal{B}_1$, and they flip their contents $\Omega_{\mathcal{B}_1}$ into $\cw$, the user collects the answer strings $\left(A_{\mathcal{B}_1}^{[i]}(\cw),A_{\mathcal{B}_2}^{[i]}(\tilde{\cw}),A_\cu^{[i]}(\tilde{\cw})\right)$.
\end{enumerate}
Since $A_\cu^{[i]}(\cw)=A_\cu^{[i]}(\tilde{\cw})$, there is no way for the user to differentiate between the two scenarios. Hence, the user commits an error either directly (if $\cw$ and $\tilde{\cw}$ differ in $W_i$) or indirectly (if $\cw$ and $\tilde{\cw}$ differ in any message other than $W_i$, as the user fails in canceling the interference from the answer strings). This is a contradiction to the reliability constraint $H(W_i|A_{1:N}^{[i]}, Q_{1:N}^{[i]})=0$.
\end{Proof}

Now, we continue with the main body of the converse proof. From Lemma~\ref{lemma:uniqueness}, the answers $A_\cu^{[i]}(\cw)$ are unique for every $\cw$, hence restricting the decoding function to these answers uniquely determine $W_i$, i.e., there exists no further confusion about the correct database contents $\mathcal{W}$, and the answering strings are designed to retrieve $W_i$ from this $\cw$. Consequently, if the true realization of the database is $\cw$, we can write
\begin{align}
R &= \frac{L}{\sum_{n=1}^N H(A_n^{[i]})} \\
  &\leq \frac{L}{\sum_{n=1}^N H(A_n^{[i]}|\cq)}\\
  &=\label{converse0}\frac{N-2B}{N}\cdot\frac{L}{(N-2B)H(A_1^{[i]}|\cq)} \\
  &=\label{converse1} \frac{N-2B}{N}\cdot \frac{L}{\sum_{n \in \cu} H(A_n^{[i]}(\cw)|\cq)}\\
  &\label{converse2}\leq \frac{N-2B}{N} \cdot C_T(|\cu|) \\
  &=\frac{N-2B}{N}\cdot C_T(N-2B)\\
  &\label{converse3}= \frac{N-2B}{N}\cdot\frac{1-\frac{T}{N-2B}}{1-\left(\frac{T}{N-2B}\right)^M}
\end{align}
where $C_T(\cdot)$ is the capacity of the PIR problem with $T$ colluding databases as a function of the number of  databases. Here, (\ref{converse0}) follows from the symmetry assumption, (\ref{converse1}) follows from the fact that $A_\cu^{[i]}(\cw)$ can decode $W_i$ correctly and then $\frac{L}{\sum_{n \in \cu} H(A_n^{[i]}(\cw)|\cq)}$ is a valid upper bound on the retrieval rate under the $T$-privacy constraint if the accessed databases are restricted to $\cu$, which is further upper bounded by the TPIR capacity $C_T(|\cu|)$ in (\ref{converse2}) as $C_T(|\cu|)$ is the supremum of all rates that can be achieved using the set of databases $\cu$ under the $T$-privacy constraint, and (\ref{converse3}) follows from the capacity expression in \cite{JafarColluding}.

\section{Conclusions and Future Directions}
In this paper, we investigated the PIR problem from $N$ replicated databases in the presence of $B$ Byzantine databases, and $T$-colluding databases from an information-theoretic perspective. We determined the exact capacity of the BPIR problem to be $C=\frac{N-2B}{N}\cdot \frac{1-\frac{T}{N-2B}}{1-(\frac{T}{N-2B})^M}$. The capacity expression shows the severe degradation in the retrieval rate in the presence of Byzantine databases. The expression shows that in order to correct the errors introduced by the adversarial databases, the system needs to have $2B$ redundant storage nodes. The retrieval rate is further penalized by the factor $\frac{N-2B}{N}$, which reflects the ignorance of the user which $N-2B$ databases are honest. The BPIR capacity converges to $C \rightarrow 1-2\gamma$ as $B,\,N \rightarrow \infty, \: B=\gamma N$, where $\gamma$ is the fraction of Byzantine databases. For large enough number of messages, the BPIR capacity approaches $C \rightarrow 1-\frac{2B+T}{N}$. We extended the optimal scheme for the RPIR problem to permit \emph{error correction} of any error pattern introduced by the Byzantine databases. The new key ingredients in the achievable scheme are: encoding the undesired messages via a punctured MDS code, successive interference cancellation to remove the interfering messages, and encoding the desired message by an outer-layer MDS code. For the converse, we adapted the cut-set bound, which was originally derived for the network coding problem against adversarial nodes, for the PIR setting.

The BPIR problem can be extended in several interesting directions. According to our formulation here, the capacities of unsynchronized and Byzantine PIR problems are the same. However, in the unsynchronized PIR problem, if the user knows in advance that at most $S$ messages are mis-synchronized, and if $S$ is small with respect to $M$, the user can potentially achieve higher rates than our formulation here, in particular, if it uses a multi-round scheme as in \cite{unsynchonizedPIR}. In addition, in modeling the mis-synchronization, if we consider some specific attack/error patterns (e.g., during mis-synchronization the stored data goes through a noisy channel with a known model), then the user can tailor an error mitigation procedure that fits these attack/error models explicitly, in contrast to our formulation here, where we assumed that the user is prepared for the worst-case errors of any structure. Finally, while we assumed that the $B$ Byzantine databases can be any one of the $\binom{N}{B}$ possible subsets, the problem can be extended to the case where only a certain subset of all possible $\binom{N}{B}$ Byzantine configurations is possible as in \cite{arbitraryCollusion} which considered a limited collusion model.

\bibliographystyle{unsrt}
\bibliography{references}
\end{document}